\documentclass[11pt,english]{article}
\pdfoutput=1
\usepackage{graphicx,array}
\usepackage{cite}
\usepackage{color}
\usepackage{latexsym}
\usepackage{amsthm}
\usepackage{amsmath}
\usepackage[titletoc]{appendix}
\usepackage{enumitem}
\usepackage{amssymb}

\usepackage[unicode=true,
 bookmarks=true,bookmarksnumbered=false,bookmarksopen=false,
 breaklinks=false,pdfborder={0 0 1},backref=false,colorlinks=true]
 {hyperref}
\hypersetup{pdftitle={Does Quantum Physics Lead To Cosmological Inflation?},
 pdfauthor={Singh and Dor\'e},
 citecolor=blue,linkcolor=blue,urlcolor=blue,citecolor=blue,linkcolor=blue}
\usepackage{breakurl}
\usepackage[hang,flushmargin]{footmisc} 

\def\simgt{\mathrel{\lower2.5pt\vbox{\lineskip=0pt\baselineskip=0pt
           \hbox{$>$}\hbox{$\sim$}}}}
\def\simlt{\mathrel{\lower2.5pt\vbox{\lineskip=0pt\baselineskip=0pt
           \hbox{$<$}\hbox{$\sim$}}}}

\newcommand{\be}{\begin{equation}}
\newcommand{\ee}{\end{equation}}
\newcommand{\bea}{\begin{eqnarray}}
\newcommand{\eea}{\end{eqnarray}}

\newcommand{\Mpl}{{ M_{pl}}}

\definecolor{nicered}{rgb}{0.7,0.1,0.1}
\definecolor{nicegreen}{rgb}{0.1,0.5,0.1}
\hypersetup{colorlinks,citecolor=black,linkcolor=black,urlcolor=black}

\definecolor{purple}{rgb}{0.5,0,0.5}
\definecolor{burgundy}{rgb}{0.5, 0.00, 0.13}

\begin{document}
\baselineskip=14pt
\hfill

\vspace{2cm}
\thispagestyle{empty}
\begin{center}
{\LARGE\bf
Does Quantum Physics Lead To Cosmological Inflation?
}\\
\bigskip\vspace{1cm}{
{\large Ashmeet Singh and Olivier Dor\'e}\footnote{e-mail: \url{ashmeet@caltech.edu, olivier.p.dore@jpl.nasa.gov}\\Corresponding Author: A.S.}
} \\[7mm]
 {\it 
Department of Physics\\
    California Institute of Technology \\
    1200 E. California Blvd., Pasadena, CA 91125\\~\\	
    
    Jet Propulsion Laboratory\\
    California Institute of Technology\\   
    4800 Oak Grove Drive, Pasadena, CA 91109  
    } \\

 \end{center}
\bigskip
\centerline{\large\bf Abstract}

\begin{quote} \small
Informed by a quantum information perspective, we interpret cosmological expansion of space as growing entanglement between underlying degrees of freedom. In particular, we focus on inflationary cosmology, which, while being a successful empirical paradigm for early universe physics, is riddled with ambiguities when one traces its quantum mechanical origins. We show, by deriving a modified cosmological continuity equation, that by properly accounting for new degrees of freedom being added to space by quantum entanglement, inflation can naturally be driven by quantum mechanics without having to resort to novel, unknown physics. 
While we explicitly focus on inflation in our discussion, we expect this approach to have possible broad implications for cosmology and quantum gravity.

\end{quote}

\vspace{3cm}


\newpage
\baselineskip=16pt
	
\setcounter{footnote}{0}

\section{Cosmology as a Quantum Gravity Lab}

The two key pillars of modern physics, quantum theory and general relativity, while being tremendously successful in making quantitative predictions for experiments, remain rather disjoint and exclusive in the scope of physical phenomena they explain. 
Only in a handful of physical setups, such as neutron stars, black holes, or cosmology, do we get a unique opportunity to study the interplay between quantum and gravitational physics. Cosmology, in particular early universe physics, offers us the ultimate quantum gravity lab, uniquely suited to study their interplay imprinted on the evolution of the cosmos as we observe it today.

On the largest scales, our universe is homogeneous and isotropic, and conforms to a classical understanding of spacetime and matter governed by Einstein's equations. To explain this, 
it is posited that the universe in its early stages underwent an inflationary phase\cite{PhysRevD.23.347,LINDE1982389}, a brief period of exponential expansion of space. Quantum fluctuations seeded in this inflationary epoch are then stretched with the expansion of space\cite{Mukhanov:1981xt,PhysRevLett.49.1110}, and eventually lead to all structure we observe today. 
In one stroke, this idea unveils a beautiful connection between the largest structures in the cosmos and the fundamental laws of physics at the smallest scales. While conceptually and empirically compelling, the exact mechanism behind inflation, the seeding of perturbations, and the ensuing quantum dynamics are not well understood. 
Inflation typically requires some form of new physics, either as new fields or dynamics, and in this effort to induce exponential expansion, the compatibility of these mechanisms with quantum physics is often unclear. 

The role of quantum physics in cosmology, albeit not yet fully understood, is expected to be even more profound, going beyond its role in seeding structure-forming perturbations. Recent advances in high energy physics and quantum information point toward a deeper connection, whereby space itself is a consequence of entanglement of underlying quantum degrees of freedom (dofs) \cite{er-eprmvr,Cao:2016mst}. In this essay, we adopt a fresh perspective on quantum cosmology by taking this ``entanglement-geometry'' connection seriously. It interprets cosmic expansion as growing entanglement between the ``qubits'' which make up space \cite{bao_etal2017}. We show, by deriving a modified cosmological continuity equation in this paradigm, that quantum physics can naturally drive inflation without the need for any new or exotic physics. While we explicitly focus on inflation in our discussion, we expect this approach to possibly have broader implications for cosmology as well.

\section{Ambiguities in Inflation as We Know It}
\label{sec:inflation_compatibility}
Inflationary dynamics, while strongly favored by cosmological observations\cite{Akrami:2018odb}, are rather non-trivial to obtain using standard constructions in both classical and quantum field theory. 
To demonstrate this, consider an expanding, homogeneous and isotropic universe in a 3+1 dimensional spacetime with a flat\cite{Akrami:2018odb} Friedmann-Robertson-Walker (FRW) metric parametrized by the scale factor $a(t)$. 
Sources of mass-energy are quantified based on their energy density $\rho$ and pressure $P$ contributions, or equivalently the equation of state $w = P/\rho$. The Hubble parameter $H = \dot{a}/a$ is governed by the Friedmann equation (in natural units with $c = \hbar = 1$, the gravitational coupling $G$ and Planck mass $\Mpl$ are connected by $8 \pi G = \Mpl^{-2}$),
\begin{equation}
\label{eq:Friedmann}
\left( \frac{\dot{a}}{a}\right)^{2} \: = \: \frac{\rho}{3 M^{2}_{pl}}  \: ,
\end{equation}
and the energy density follows the general relativistic continuity equation,
\be
\label{eq:continuity_GR}
\dot{\rho} \: = \: -3 H \left(\rho + P\right) \: ,
\ee
which, for a constant $w$ (as for the simplest models), yields $\rho(a) \propto a^{-3(1 + w)}$. 
In the usual cosmology folklore, the universe begins with an inflationary epoch with exponential growth characterized by $w = P/\rho = -1$ (or equivalently $\rho = \mathrm{constant}$), followed by radiation domination with $w = 1/3$, then a matter-dominated phase with $w = 0$, and finally, late time evolution is dominated by a cosmological constant-like contribution having $w = -1$. 
\\
\\
Deriving $w = -1$ from conventional sources of energy-momentum (either classical or quantum) is not easy to come by, as we will now discuss.
\\
\\
In classical vanilla cosmology, one typically focuses on ``slow roll'' models of inflation\cite{mukhanov_2005}, which posit a \emph{new} scalar field, the ``inflaton'' that slowly evolves on a relatively flat potential energy curve. With the right initial conditions and large scale homogeneity, the potential energy contribution dominates in both $\rho$ and $P$, albeit with opposite signs, implying $P = -\rho$, or equivalently $w = -1$. 
This is, however, a strongly classical understanding of scalar field dynamics where both the field and its conjugate momentum are known, in violation of the Heisenberg uncertainty principle. Even if one predicates this picture to apply for quantum mechanical expectation values, it is unclear as to how to ab-initio assign a quantum state to the relevant dofs. Not only is the dynamics contrived, its compatibility with quantum mechanics is loose at best.

On the other hand, one can try and directly source inflationary dynamics from quantum field theory (QFT), by looking at energy density of \emph{vacuum} QFT modes. In this approach, even before the question of whether new fields are required, we find an inconsistency in the time dependence of the vacuum energy. Contrary to a-priori expectations, QFT based calculations\cite{Martin:2012bt} do not yield the desired/expected $w = -1$ corresponding to the vacuum\footnote{This inconsistency can be tied to the discussion of broken Lorentz invariance in the setup when one assumes a non-Lorentz covariant ultraviolet (UV) cutoff, and efforts to fix it using alternative regularization schemes give pathologies such as a negative energy density\cite{Martin:2012bt,Mathur:2020ivc}.}. For example, for the simplest case of a free scalar Klein-Gordon field, 
the energy eigenvalues, including the vacuum energy (for each Fourier $\vec{k}$ mode, and therefore the total as well), follow,
\be 
\label{eq:KG_eigen}
\rho_{\vec{k}} \: \propto \: \: \frac{1}{a^{3}} \sqrt{\left(\frac{k}{a} \right)^{2} + m^{2} }  \: \: \propto \begin{cases}
\frac{1}{a^{4}} \: , \: \: \: \: \frac{k}{a} >> m \\
\\
\frac{1}{a^{3}} \: , \: \: \: \:\frac{k}{a} << m 
\end{cases},
\ee
corresponding to either radiation-like ($\rho \propto 1/a^{4} \Longleftrightarrow w = 1/3$) or matter-like ($\rho \propto 1/a^{3} \Longleftrightarrow w = 0$) equation of state, but \emph{not that of inflation} which needs $\rho = \mathrm{constant}$.
\\
\\
Exponential expansion sourced by a constant energy density therefore seems elusive! We will now discuss how inflation can naturally occur by treating expansion of space from a quantum mechanical viewpoint.
\section{Inflation from Quantum Mechanics}
\begin{itemize}
\item \textbf{\large{Expansion of Space from Quantum Entanglement}}
\end{itemize}
Classically, spacetime is a given entity, treated as a smooth metric manifold, which supports wavelengths both arbitrarily large and small. In an expanding FRW geometry, these modes stretch with the Hubble flow simply as a consequence of the changing metric distance, without reference to any underlying dofs which make up space. On the other hand, from a quantum information perspective, space \emph{emerges} from quantum entanglement\cite{vanRaamsdonk2010}. In particular, we focus on the ``Quantum Circuit Cosmology'' (QCC) picture\cite{bao_etal2017, Bao:2017qmt} which 
interprets expansion of the space as a quantum circuit, progressively entangling more dofs (corresponding to larger space, and matter fields on this background) as time evolves. To be able to do this consistently, it leverages a key implication of black hole thermodynamics and holography which suggests that there are a finite number of dofs in a given region of space\cite{Bao:2017rnv}.

The QCC posits that the universe begins (presumably with size $L_{ini} \sim \Mpl^{-1}$, before which a classical metric ceases to be well defined) with only a few dofs entangled.
New ``ancilla'' (initially unentangled) dofs get entangled with time to increase the size of the universe, thereby giving an infrared (IR) size, $L(t)$,
\begin{equation}
L(t)  = \frac{a(t)}{a_{ini}} L_{ini}  \: \equiv \: a(t) L_{0} \: ,
\end{equation}
where $a_{ini}$ is the initial scale factor, and $L_{0} \equiv L_{ini}/a_{ini}$.
While the IR size (generally different from the particle or Hubble horizons) grows with $a(t)$, new dofs are continually added by entanglement in the ultraviolet, keeping the UV scale fixed\footnote{This idea is closely linked to the discussion of broken Lorentz invariance in Mathur's recent work\cite{Mathur:2020ivc}. Our discussion approaches the question from a somewhat different perspective and the conclusions were obtained independently of those by Mathur.} \footnote{It is also worth mentioning that in the QCC picture, one sidesteps the transplanckian problem\cite{bao_etal2017} by never referring to the existence of a sea of high energy excitation modes frozen beyond the Planck scale which get dynamic as they expand with the Hubble flow. Instead, modes are added at a fixed UV scale as a consequence of entanglement.} at $\Lambda$, presumably around the Planck scale $\Lambda \sim \Mpl$ \cite{Mathur:2020ivc}.

\begin{itemize}
\item \textbf{\large{New modes modify the Universe expansion history}}
\end{itemize}
The key aspect of QCC is the \emph{introduction of new dofs}, both which make up spacetime \emph{and matter fields} on this background.
As a consequence, the set of allowed QFT modes increases in time. More concretely, in comoving coordinates \big(recall that a comoving mode $\vec{k}_{c}$ is related to the physical mode $\vec{k}$ by $\vec{k}_{c} = a(t)\vec{k}$\big), the allowed discrete set $\mathbb{K}_{c}(t)$ of comoving Fourier modes is therefore given by the following set,
\begin{equation}
\mathbb{K}_{c}(t) = \bigg\{ \vec{k}_{c} \equiv (k_{x}, k_{y}, k_{z}) \: | \: k_{j} = \frac{1}{L_{0}}, \frac{2}{L_0} ,\ldots , a(t)\Lambda \bigg\} \: .
\end{equation}
Physically, this implies that the most IR mode keeps getting redshifted as $L^{-1}(t)$, while keeping the UV scale fixed at $\Lambda$, or equivalently from a comoving perspective, the IR mode remains fixed at $L^{-1}_{0}$, while new modes are added deeper in the UV with $\Lambda_c(t) = a(t)\Lambda$. 
The UV and IR scales are much closer in the beginning, and progressively get further separated as time evolves.

Whereas in a classical understanding, 
modes exist by fiat and only stretch with the Hubble flow, in a quantum spacetime, there is an addition of new modes which contribute to the energy-momentum,
\be
\rho \:  \: \: = \: \sum_{\vec{k} \in \mathbb{K}_{c}(t)}^{\Lambda_{c} = a(t)\Lambda} \rho_{\vec{k}} \: .
\ee
Each mode already ``initialized'' as part of the entanglement-based expansion of space still satisfies the usual continuity equation,
\be
\dot{\rho}_{\vec{k}} \: = \: -3 H \left(\rho_{\vec{k}} + P_{\vec{k}} \right) \: ,
\ee
stretching and/or diluting with expansion. However, as a consequence of addition of new modes (and mathematically of the Leibnitz integration rule), we get a \emph{modified continuity equation}, 
\be
\label{eq:continuity_QCC}
\dot{\rho} \: = \: -3 H \left(\rho + P\right) + 3 H a^{3}(t) \Lambda^{3} L^{3}_{0}  \: \rho_{\Lambda_{c}(t)}    \: .
\ee

\begin{itemize}
\item \textbf{\large{Inflation with $w\ne -1$}}
\end{itemize}

While the standard continuity equation of  Eq. (\ref{eq:continuity_GR}) typically only allows $\rho$ to decrease with expansion, the new QCC term $\big(3 H a^{3}(t) \Lambda^{3} L^{3}_{0}  \: \rho_{\Lambda_{c}(t)} \big)$ being positive can offset the decrease to give $\rho = \mathrm{constant}$, corresponding to inflation. Therefore, $w = -1$ is \emph{not} the only way to obtain exponential expansion of space. Indeed, a simple calculation with a generic Klein-Gordon field shows that even though QFT modes satisfy $ 0 \leq w \leq 1/3$ (as showed in section \ref{sec:inflation_compatibility}), the total energy density \emph{does} remain constant, leading to inflationary dynamics, 
\be
\label{eq:KG_const_rho1}
\rho \:  \: \: = \: \sum_{\vec{k} \in \mathbb{K}_{c}(t)}^{\Lambda_{c} = a(t)\Lambda} \rho_{\vec{k}} \: \:  \approx \: \:  \frac{1}{a^3}\int^{a(t)\Lambda} \:  4 \pi k^{2} \: \Bigg( \sqrt{\left(\frac{k}{a} \right)^{2} + m^{2} } \Bigg) \: \:   dk \: \: \sim \:  \: \Lambda^4 \:  \: = \: \mathrm{constant} \: .
\ee
Thus, a consequence of interpreting cosmological expansion as growing quantum entanglement naturally lends inflation a more rigorous quantum mechanical footing without the need for new fields or exotic dynamics.
\\
\\
The implications of this approach will presumably go beyond inflation, non-trivially impacting the dynamics of cosmological evolution. It will be an interesting extension to study how a quantum-first approach, in a unified paradigm, can seed perturbations due to evolution of the quantum states of different QFT modes, and how it leads to exit from inflation into other epochs. 

\begin{center} 
{\bf Acknowledgments}
\end{center}

\textcopyright 2021. We would like to thank ChunJun (Charles) Cao and Swati Chaudhary for helpful discussions. We acknowledge the generous support of the Heising-Simons Foundation. Part of the research described in this paper was carried out at the Jet Propulsion Laboratory, California Institute of Technology, under a contract with the National Aeronautics and Space Administration.

\bibliographystyle{utphys}
\bibliography{quantum_frw_essay}

\providecommand{\href}[2]{#2}\begingroup\raggedright\begin{thebibliography}{10}

\bibitem{PhysRevD.23.347}
A.~H. Guth, ``Inflationary universe: A possible solution to the horizon and
  flatness problems,'' \href{http://dx.doi.org/10.1103/PhysRevD.23.347}{{\em
  Phys. Rev. D} {\bfseries 23} (Jan, 1981) 347--356}.
  \url{https://link.aps.org/doi/10.1103/PhysRevD.23.347}.

\bibitem{LINDE1982389}
A.~Linde, ``A new inflationary universe scenario: A possible solution of the
  horizon, flatness, homogeneity, isotropy and primordial monopole problems,''
  \href{http://dx.doi.org/https://doi.org/10.1016/0370-2693(82)91219-9}{{\em
  Physics Letters B} {\bfseries 108} no.~6, (1982) 389--393}.
  \url{https://www.sciencedirect.com/science/article/pii/0370269382912199}.

\bibitem{Mukhanov:1981xt}
V.~F. Mukhanov and G.~V. Chibisov, ``{Quantum Fluctuations and a Nonsingular
  Universe},'' {\em JETP Lett.} {\bfseries 33} (1981) 532--535.

\bibitem{PhysRevLett.49.1110}
A.~H. Guth and S.-Y. Pi, ``Fluctuations in the new inflationary universe,''
  \href{http://dx.doi.org/10.1103/PhysRevLett.49.1110}{{\em Phys. Rev. Lett.}
  {\bfseries 49} (Oct, 1982) 1110--1113}.
  \url{https://link.aps.org/doi/10.1103/PhysRevLett.49.1110}.

\bibitem{er-eprmvr}
M.~{van Raamsdonk}, ``{Building up spacetime with quantum entanglement},''
  \href{http://dx.doi.org/10.1007/s10714-010-1034-0}{{\em General Relativity
  and Gravitation} {\bfseries 42} (Oct., 2010) 2323--2329},
  \href{http://arxiv.org/abs/1005.3035}{{\ttfamily arXiv:1005.3035 [hep-th]}}.

\bibitem{Cao:2016mst}
C.~Cao, S.~M. Carroll, and S.~Michalakis, ``{Space from Hilbert Space:
  Recovering Geometry from Bulk Entanglement},''
  \href{http://dx.doi.org/10.1103/PhysRevD.95.024031}{{\em Phys. Rev.}
  {\bfseries D95} no.~2, (2017) 024031},
\href{http://arxiv.org/abs/1606.08444}{{\ttfamily arXiv:1606.08444 [hep-th]}}.

\bibitem{bao_etal2017}
N.~Bao, C.~Cao, S.~M. Carroll, and L.~McAllister, ``{Quantum Circuit Cosmology:
  The Expansion of the Universe Since the First Qubit},''
\href{http://arxiv.org/abs/1702.06959}{{\ttfamily arXiv:1702.06959 [hep-th]}}.

\bibitem{Akrami:2018odb}
{\bfseries Planck} Collaboration, Y.~Akrami {\em et~al.}, ``{Planck 2018
  results. X. Constraints on inflation},''
  \href{http://dx.doi.org/10.1051/0004-6361/201833887}{{\em Astron. Astrophys.}
  {\bfseries 641} (2020) A10},
  \href{http://arxiv.org/abs/1807.06211}{{\ttfamily arXiv:1807.06211
  [astro-ph.CO]}}.

\bibitem{mukhanov_2005}
V.~Mukhanov, \href{http://dx.doi.org/10.1017/CBO9780511790553}{{\em Physical
  Foundations of Cosmology}}.
\newblock Cambridge University Press, 2005.

\bibitem{Martin:2012bt}
J.~Martin, ``{Everything You Always Wanted To Know About The Cosmological
  Constant Problem (But Were Afraid To Ask)},''
  \href{http://dx.doi.org/10.1016/j.crhy.2012.04.008}{{\em Comptes Rendus
  Physique} {\bfseries 13} (2012) 566--665},
  \href{http://arxiv.org/abs/1205.3365}{{\ttfamily arXiv:1205.3365
  [astro-ph.CO]}}.

\bibitem{Mathur:2020ivc}
S.~D. Mathur, ``{Three puzzles in cosmology},''
  \href{http://dx.doi.org/10.1142/S021827182030013X}{{\em Int. J. Mod. Phys. D}
  {\bfseries 29} no.~14, (2020) 2030013},
  \href{http://arxiv.org/abs/2009.09832}{{\ttfamily arXiv:2009.09832
  [hep-th]}}.

\bibitem{vanRaamsdonk2010}
M.~Van~Raamsdonk, ``{Building up spacetime with quantum entanglement},''
  \href{http://dx.doi.org/10.1007/s10714-010-1034-0,
  10.1142/S0218271810018529}{{\em Gen. Rel. Grav.} {\bfseries 42} (2010)
  2323--2329}, \href{http://arxiv.org/abs/1005.3035}{{\ttfamily arXiv:1005.3035
  [hep-th]}}.
[Int. J. Mod. Phys.D19,2429(2010)].

\bibitem{Bao:2017qmt}
N.~Bao, C.~Cao, S.~M. Carroll, and A.~Chatwin-Davies, ``{De Sitter Space as a
  Tensor Network: Cosmic No-Hair, Complementarity, and Complexity},''
  \href{http://dx.doi.org/10.1103/PhysRevD.96.123536}{{\em Phys. Rev. D}
  {\bfseries 96} no.~12, (2017) 123536},
  \href{http://arxiv.org/abs/1709.03513}{{\ttfamily arXiv:1709.03513
  [hep-th]}}.

\bibitem{Bao:2017rnv}
N.~Bao, S.~M. Carroll, and A.~Singh, ``{The Hilbert Space of Quantum Gravity Is
  Locally Finite-Dimensional},''
  \href{http://dx.doi.org/10.1142/S0218271817430131}{{\em Int. J. Mod. Phys.}
  {\bfseries D26} no.~12, (2017) 1743013},
\href{http://arxiv.org/abs/1704.00066}{{\ttfamily arXiv:1704.00066}}.

\end{thebibliography}\endgroup

\end{document}